\documentclass[fleqn,usenatbib,useAMS]{mnras}

\usepackage{mathptmx}

\usepackage[T1]{fontenc}
\usepackage{ae,aecompl}

\usepackage{graphicx}	
\usepackage{amsmath}	
\usepackage{amssymb}	
\usepackage{multicol}
\usepackage[export]{adjustbox}

\newcommand{\pivec}{\mbox{\boldmath $\pi$}}

\usepackage{ulem}

\newcommand{\thS}{\theta_*}

\title[Stellar binary microlensing in the Galactic bulge: OGLE-2015-BLG-0060]{An analysis of binary microlensing event OGLE-2015-BLG-0060}

\author[Y. Tsapras et al.]{Y. Tsapras$^{1}$\thanks{E-mail: ytsapras@ari.uni-heidelberg.de}
A. Cassan$^{2}$,
C. Ranc$^{M3}$,
E. Bachelet$^{3}$,
R. Street$^{3}$,
A. Udalski$^{O1}$,
\newauthor
M. Hundertmark$^{1}$,
V. Bozza$^{R10,S2}$,
J. P. Beaulieu$^{2,4}$,
J. B Marquette$^{2}$,
E. Euteneuer$^{1}$,
\newauthor
{\it The RoboNet team:}
D. M. Bramich$^{R18}$,
M. Dominik$^{R3}$,
R. Figuera Jaimes$^{R3,R16}$,
\newauthor
K. Horne$^{R3}$,
S. Mao$^{R16,R15,R17}$,
J. Menzies$^{R8}$ 
R. Schmidt$^{1}$,
C. Snodgrass$^{R13,R19}$,
\newauthor
I. A. Steele$^{R4}$,
J. Wambsganss$^{1,R20}$
\newauthor
{\it The OGLE collaboration:}
P. Mr{\'o}z$^{O1}$,
M.K. Szyma{\'n}ski$^{O1}$,
I. Soszy{\'n}ski$^{O1}$,
J. Skowron$^{O1}$,
\newauthor
P. Pietrukowicz$^{O1}$
S. Koz{\l}owski$^{O1}$,
R. Poleski$^{O2}$,
K. Ulaczyk$^{O1}$,
M. Pawlak$^{O1}$ 
\newauthor
{\it The MiNDSTEp collaboration:}
U. G. J{\o}rgensen$^{P1}$,
J. Skottfelt$^{P23,P1}$,
A. Popovas$^{P1}$,
\newauthor
S. Ciceri$^{P10}$,
H. Korhonen$^{P21,P11,P1}$,
M. Kuffmeier$^{P1}$,
D. F. Evans$^{P14}$,
N. Peixinho$^{P15, P27}$,
\newauthor
T. C. Hinse$^{P16}$,
M. J. Burgdorf$^{P17}$,
J. Southworth$^{P14}$,
R. Tronsgaard$^{P18}$,
E. Kerins$^{R17}$,
\newauthor
M.I. Andersen$^{P21}$,
S. Rahvar$^{P22}$,
Y. Wang$^{P24}$,
O. Wertz$^{P25}$,
M. Rabus$^{P26,P10}$,
\newauthor
S. Calchi Novati$^{S3}$,
G. D'Ago$^{S4,R10}$,
G. Scarpetta$^{S5,R10}$,
L. Mancini$^{P28,P29,P30}$
\newauthor
{\it The MOA collaboration:}
F. Abe$^{M2}$,
Y. Asakura$^{M2}$,
D. P.~Bennett$^{M3,M12}$,
\newauthor
A. Bhattacharya$^{M3,M12}$,
M. Donachie$^{M4}$,
P. Evans$^{M4}$,
A. Fukui$^{M5}$,
Y. Hirao$^{M1}$,
\newauthor
Y. Itow$^{M2}$,
K. Kawasaki$^{M1}$,
N. Koshimoto$^{M1}$,
M.C.A. Li$^{M4}$,
C.H. Ling$^{M6}$,
K. Masuda$^{M2}$,
\newauthor
Y. Matsubara$^{M2}$,
Y. Muraki$^{M2}$,
S. Miyazaki$^{M1}$,
M. Nagakane$^{M1}$,
K. Ohnishi$^{M7}$,
\newauthor
N. Rattenbury$^{M4}$,
To. Saito$^{M8}$,
A. Sharan$^{M4}$,
H. Shibai$^{M1}$,
D.J. Sullivan$^{M9}$,
\newauthor
T. Sumi$^{M1}$,
D. Suzuki$^{M3}$,
P.J. Tristram$^{M10}$,
T. Yamada$^{M11}$,
A. Yonehara$^{M11}$
\\
{\it \small Affiliations appear at the end of the paper}\\
}

\date{Published in MNRAS 30 May 2019. \href{https://doi.org/10.1093/mnras/stz1404}{https://doi.org/10.1093/mnras/stz1404}}

\pubyear{2019}

\begin{document}
\label{firstpage}
\pagerange{\pageref{firstpage}--\pageref{lastpage}}
\maketitle

\begin{abstract}
    We present the analysis of stellar binary microlensing event OGLE-2015-BLG-0060 based on observations obtained from 13 different telescopes. Intensive coverage of the anomalous parts of the light curve was achieved by automated follow-up observations from the robotic telescopes of the Las Cumbres Observatory. We show that, for the first time, all main features of an anomalous microlensing event are well covered by follow-up data, allowing us to estimate the physical parameters of the lens. The strong detection of second-order effects in the event light curve necessitates the inclusion of longer-baseline survey data in order to constrain the parallax vector. We find that the event was most likely caused by a stellar binary-lens with masses $M_{\star1} = 0.87 \pm 0.12 M_{\odot}$ and $M_{\star2} = 0.77 \pm 0.11 M_{\odot}$. The distance to the lensing system is 6.41 $\pm 0.14$ kpc and the projected separation between the two components is 13.85 $\pm 0.16$ AU. Alternative interpretations are also considered.
\end{abstract}




\section{Introduction}
    There are two unique aspects to gravitational microlensing that set it apart from other exoplanet detection methods. Firstly, it is most sensitive to planets at separations $\sim$1-10AU from their hosts \citep{Tsapras2016,suzuki2016,cassan2012,Tsapras2003}, a region of great relevance to planetary formation theories \citep{ida2013}, as this typically places the planet beyond the {\it snow-line}\footnote{The {\it snow-line} is the distance from a proto-star beyond which any water present in the proto-planetary disk will be in the form of ice grains.} of the host star \citep{armitage2016}, a region largely inaccessible to the transit and radial-velocity methods. Secondly, it detects planets around faint stars at distances of several thousand parsec \citep{penny2016}, whereas almost every planet discovered to date by other methods lies only within a few hundred parsec from the Sun. Since stellar metallicity decreases with distance from the centre of the Galaxy \citep{ivezic2008}, microlensing planets may have formed in more metal-rich environments leading to a potentially different statistical distribution compared to the sample of nearby planets. This hypothesis can only be explored through microlensing.

    The phenomenon of gravitational microlensing occurs when a foreground star gravitationally lenses a luminous background star, causing its brightness to gradually increase, and then gradually decrease, over a period of several days to months \citep{paczynski1986}. The angular distance between the images generated by the lensing event is generally too small to resolve with current technology so that only the change in brightness of the {\it background} object, commonly referred to as the {\it source} star, is observed. The only known exception to date is the nearby microlensing event TCP J0507+2447, detected on October 25$^{\mathrm{th}}$ 2017 by Japanese amateur astronomer Tadashi Kojima, for which \citet{2019ApJ...871...70D} managed to resolve, for the first time, the two microlensing images using the GRAVITY interferometer on the Very Large Telescope (VLT). A few very bright microlensing events per year might also be within the reach of the PIONIER instrument \citep{2016MNRAS.458.2074C}.
    
    Should the {\it foreground} lensing object be a star-star or star-planet system, its exact geometric alignment and physical parameters may leave an imprint on an otherwise symmetric light curve \citep{dominik2010,gaudi2012,tsapras2018}. These anomalous features can be detected and sampled with frequent ($\sim$hourly to daily) observations, depending on the particular event. Due to the transient and unpredictable nature of microlensing events, it is often not possible to distinguish between planetary and stellar binary anomalies when they are first identified. Follow-up observations are therefore typically executed for almost all events where there is evidence of ongoing anomalies and the light curves are continuously re-assessed through real-time modeling. A full characterisation is usually obtained only after the event has expired and returned to its baseline brightness. 

    Recent results from microlensing campaigns suggest that ice and gas giant planets are a relatively common feature ($\sim$35\%) around K and M-dwarf stars \citep{Gould2010}. Microlensing searches have also identified a number of very massive cool planets \citep{batista2011,tsapras2014,koshimoto2014,skowron2015} and brown dwarf companions around low-mass stars \citep{street2013,ranc2015,han2016}, as well as several terrestrial to sub-Neptune mass planets \citep{beaulieu2006,kubas2012,gould2014,shvartzvald2017}, systems with multiple planets \citep{gaudi2008,han2013} and the first possible detection of an exomoon \citep{bennett2014}.

    In this paper we present the analysis of binary microlensing event OGLE-2015-BLG-0060 using observations collected from 13 different telescopes spread out around the world, providing continuous monitoring of the event. Stellar binary microlensing events are discovered far more often than planetary ones and their diverse morphologies have been the subject of several past studies \citep{shin2017,han2016b,shin2012,jaroszynski2010,skowron2007}. Typically, their caustic-crossing features are predicted well in advance, given differences in the smoothly rising part of the light curve as compared to single-lens events. OGLE-2015-BLG-0060 is of particular interest because it is the first time automated follow-up observations have achieved excellent coverage of all anomalous features without human involvement, demonstrating the potential of fully robotic observations in characterising microlensing events. 
    
    In Section \ref{sec:obs} we provide a summary of the observations and data analysis. Section \ref{sec:mod} describes the steps taken to model the event light curve and Section \ref{sec:phy} the method used to determine the physical properties of the system. Finally, we provide a summary and conclusions in Section \ref{sec:con}.

\section{Observations and photometry}
\label{sec:obs} 
    \begin{table*}
	    \centering
    	\caption{Data sets used in this analysis and their properties}
	    \label{tab:obs}
	    \begin{tabular}{llcl} 
		    \hline
		    \hline
    	    Group & Telescope & Passband & Data points\\
		    \hline
		    OGLE & 1.3m Warsaw Telescope, Las Campanas, Chile & $I,V$ & 1629,101\\
		    MOA & 1.8m MOA Telescope, Mount John, New Zealand & $MOAred,MOAblue$ & 5674,71\\
        	RoboNet & 1.0m LCO (Dome A), CTIO, Chile & SDSS-i' & 176\\
	    	RoboNet & 1.0m LCO (Dome B), CTIO, Chile & SDSS-i' & 83\\
	    	RoboNet & 1.0m LCO (Dome C), CTIO, Chile & SDSS-i' & 138\\
	    	RoboNet & 1.0m LCO (Dome A), SAAO, South Africa & SDSS-i' & 92\\		
	    	RoboNet & 1.0m LCO (Dome B), SAAO, South Africa & SDSS-i' & 60\\
	    	RoboNet & 1.0m LCO (Dome C), SAAO, South Africa & SDSS-i' & 86\\
	    	RoboNet & 1.0m LCO (Dome A), SSO, Australia & SDSS-i'& 65\\
	    	RoboNet & 1.0m LCO (Dome B), SSO, Australia & SDSS-i' & 62\\		
    		MiNDSTEp & 1.5m Danish Telescope, La Silla, Chile & LIred & 85\\		
    		MiNDSTEp & 0.6m Salerno Telescope, Salerno, Italy & $I$ & 25\\
	    	VVV & 4.1m VISTA Telescope, Paranal, Chile & $K$ & 240\\		
		    \hline
		    \multicolumn{4}{l}{Note: For a description of the MiNDSTEp LIred bandpass see \citet{skottfelt2015}.} \\
		    \multicolumn{4}{l}{For a description of the MOA $MOAred,MOAblue$ bandpasses see \citet{2008ExA....22...51S}.}
    	\end{tabular}
    \end{table*}

\subsection{Survey and follow-up observations}
    Microlensing event OGLE-2015-BLG-0060 was announced on 2015 February 17 by the Early Warning System (EWS)\footnote{http://ogle.astrouw.edu.pl/ogle4/ews/ews.html} of the Optical Gravitational Lensing Experiment (OGLE) survey \citep{Udalski2003,Udalski2015} at equatorial coordinates $\alpha = 17\textsuperscript{h}59\textsuperscript{m}58.35\textsuperscript{s}$, $\delta = -27^{\circ}46\arcmin51.4\arcsec$ (J2000) ($l,b = 2.6005^{\circ}, -2.1315^{\circ}$). OGLE observations were carried out with the 1.3-m Warsaw telescope at Las Campanas Observatory in Chile, with the 32-chip mosaic CCD camera. The event occurred in OGLE bulge field 504, which was imaged several times per night when not interrupted by weather or the full Moon, providing good coverage of the light curve when the bulge was visible from Chile. The OGLE survey reported a baseline $I_0$-band magnitude for the blended star of 16.683, which was later revised to 16.933. The predicted maximum magnification at the time of announcement was very low, therefore the target was originally considered low priority for follow-up observations. The event was also independently picked up by the MOA survey \citep{sumi2003}, using the 1.8-m MOA survey telescope at Mount John observatory in New Zealand, on 2015 March 17, and designated MOA-2015-BLG-071. 

    \begin{figure*}
	    \includegraphics[width=2\columnwidth]{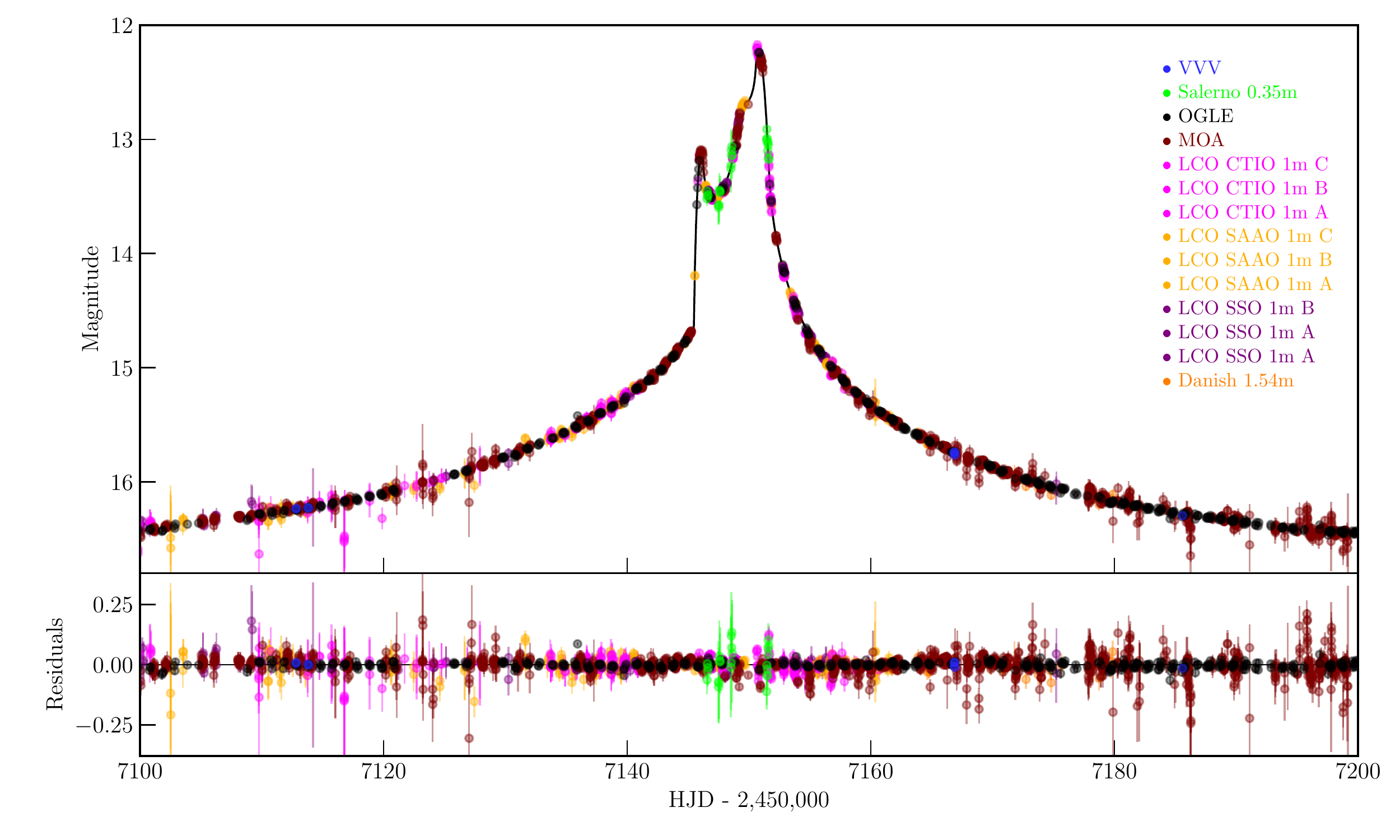}
        \caption{Light curve of microlensing event OGLE-2015-BLG-0060 showing the best-fit binary model including parallax. The legend on the right of the figure lists the contributing telescopes.}
        \label{fig:lightcurve}
    \end{figure*}

    By mid-March 2015 it became apparent that it might be a high magnification event and the RoboNet \citep{Tsapras2009} and MiNDSTEp \citep{dominik2010b} teams began to observe it automatically. RoboNet observations were carried out using the southern ring of 1m robotic telescopes of the Las Cumbres Observatory (LCO) \citep{brown2013}, a total of 8 telescopes located at the Cerro Tololo International Observatory (CTIO) in Chile, South African Astronomical Observatory (SAAO) in South Africa and Siding Spring Observatory (SSO) in Australia, providing continuous coverage of the event light curve. MiNDSTEp observations were carried out on the Danish 1.54m telescope at ESO La Silla in Chile and the 0.6m telescope at Salerno Observatory in Italy. New photometric reductions of VVV survey observations \citep{2010NewA...15..433M} in the K-band at the location of the target were also included in the analysis, although none were obtained during the peak of the event.

    The event was also observed by the Korea Microlensing Telescopes Network (KMTNet) \citep{kim2016}, and their data are analysed separately. The Spitzer satellite also obtained observations of this target from June 12 (HJD$\sim$2457186) to July 19 (HJD$\sim$2457223), as part of an effort to constrain the parallax \citep{yee2015}, but unlike the case of OGLE-2015-BLG-0966 \citep{street2016}, these only cover the part of the light curve when the event is returning to baseline and provide no additional constraints. The Spitzer data are therefore not included in the final model the event.

    The full light curve of OGLE-2015-BLG-0060 is shown in Figure \ref{fig:lightcurve}, together with the best-fit model (cf. Section \ref{sec:mod}).

\subsection{Detection of the anomaly}
\label{detanom}
    The first evidence that the event deviated from the standard single-lens Paczy\'{n}ski curve appeared on March 11, when the SIGNALMEN anomaly detector \citep{dominik2007} triggered an alert on the ARTEMiS event monitoring system \citep{dominik2008}. The alert was propagated to the real-time modelling software RTModel \citep{bozza2010} to generate the first set of models, and complementary observation requests aimed at characterising the anomalous feature were automatically submitted to the LCO robotic telescopes \citep{hundertmark2018}. On the same day, an email by V. Bozza alerted the community that the broad peak observed at the end of the 2014 season (HJD$\sim$2456920, see Figure~\ref{fig:bump} in the Appendix) and the subsequent rise during 2015 (HJD$\sim$2457080) were incompatible with a single lens, and that binary and planetary solutions were possible\footnote{The small deviation that triggered the follow-up observations was generated by the source approaching the first set of caustics at a distance of $\sim1.1$ Einstein radii and took place well before the crossing of the second set of caustics.}. A second report by Bozza on April 20, using updated data, predicted a possible caustic crossing. On May 3 (HJD$\sim$2457145.5), strong deviations indicating a caustic crossing were detected in the RoboNet and then the OGLE and MiNDSTEp data. Shortly thereafter, D. Bennett alerted the community that a strong anomaly was ongoing, and V. Bozza pointed out that higher-order effects needed to be taken into account during the model fitting process. By May 9 (HJD$\sim$2457151.5), V. Bozza had determined that the lens was likely to be a stellar binary, and on May 19 (HJD$\sim$2457161.5) the Chungbuk National University group (CBNU, C. Han), using private KMTNet data that covered the anomalous feature, gave a preliminary estimate of the parameters of the system while the event was still ongoing. This solution was confirmed independently by V. Bozza on the same day.
    
\subsection{Data reduction}
    The data used for the analysis presented in this paper and the telescopes used for the observations are listed in Table \ref{tab:obs}. Most observations were obtained in the $I$ band (SDSS-i$^\prime$), while some images obtained by OGLE in $V$ were also used to generate a colour-magnitude diagram and classify the source star. MOA observations were performed with the MOA wide-band red filter, which is specific to that survey. We note that there are also observations obtained privately by the KMTNet survey, which will be analysed in a separate paper. 

    The photometric analysis of crowded-field observations is a challenging task. Images of the Galactic bulge contain thousands of stars whose point-spread functions (PSFs) often overlap, therefore aperture and PSF-fitting photometry offer very limited sensitivity to photometric deviations generated by the presence of low-mass planetary companions. For this reason, observers of microlensing events routinely perform difference image analysis (DIA) \citep{alard98}, which offers superior photometric sensitivity under such conditions. For a given telescope and camera, the technique of difference image analysis uses a {\it reference} image\footnote{This can be either a single image or a combination of images taken under the best seeing conditions} to which background, astrometric, photometric and point-spread-function corrections are applied to match the images of that same field taken at each individual epoch. The fitted model based on the {\it reference} image is then subtracted from the matching images to produce residual (or {\it difference} images). Stars that did not vary in brightness between the times the images were obtained leave no systematic residuals on the {\it difference} images, but stars that underwent brightness variations leave clear positive or negative residuals. 

    Most microlensing teams have developed custom DIA pipelines to reduce their observations. OGLE and MOA images were reduced using the photometric pipelines described in \citet{Udalski2003} and \citet{bond2001} respectively. RoboNet and MiNDSTEp observations were processed using customised versions of the DanDIA pipeline \citep{bramich2013,bramich2008}. Salerno data were reduced with a locally developed PSF-fitting pipeline. The data sets presented in this paper have been reprocessed to optimise photometric precision and it is these data we used as input when modelling the microlensing event. They are available for download from the online version of the paper. 

    \begin{figure*}
        \begin{center}
	        \includegraphics[width=2\columnwidth]{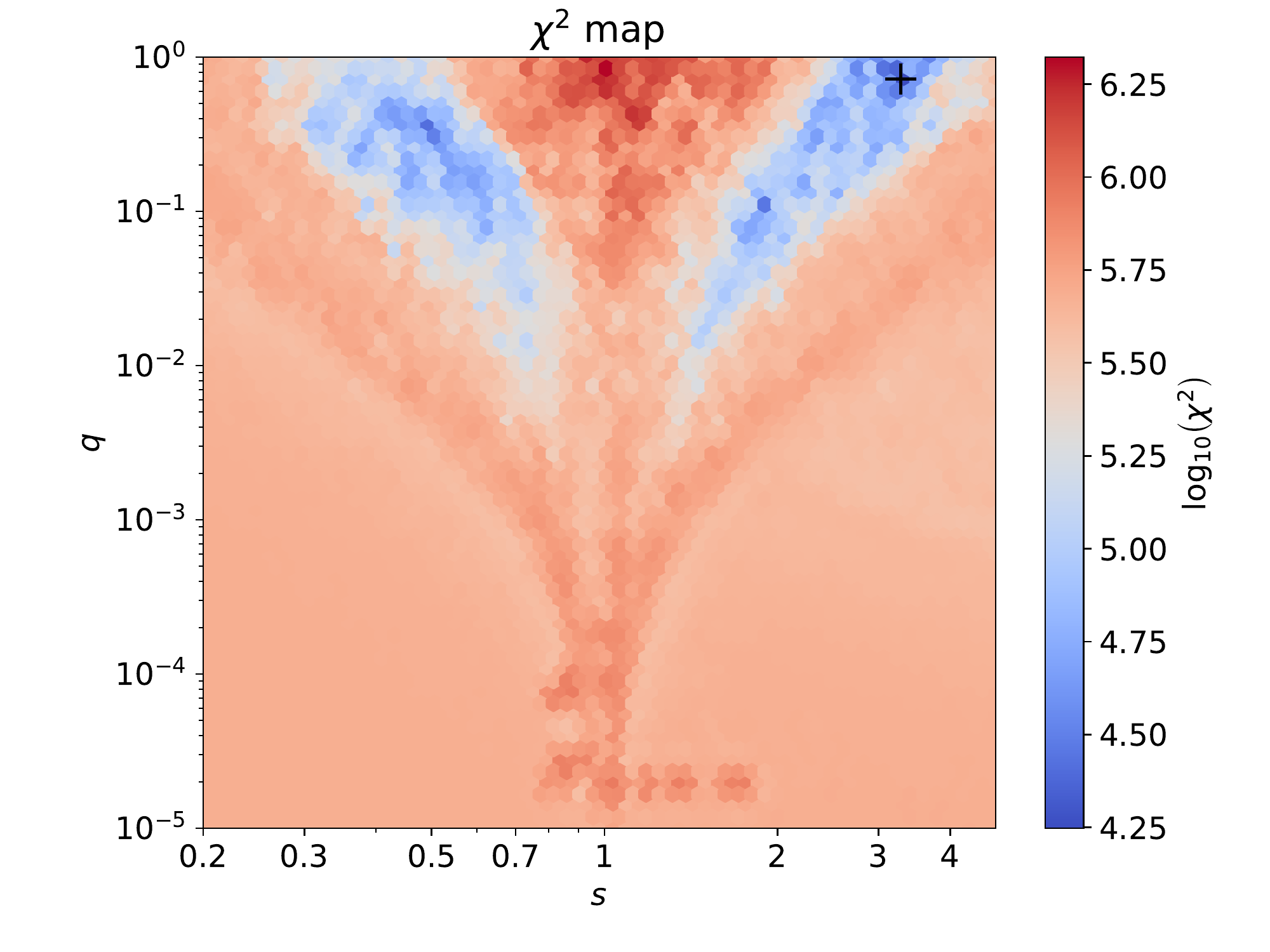}
            \caption{$\chi^2$ map as a function of $(s, q)$, assuming a static binary-lens model including finite-source effects. The underlying $s, q$-grid has 3024 grid points and spans binary-lens separations from $s=0.2$ to $s=5$ (84 values) and mass ratios from $q=10^{-5}$ to $q=1$ (36 values), both uniformly spaced in logarithmic scale. Red to blue colors are decreasing values of $\chi^2$, on a logarithmic scale displayed on the right. Two wide and symmetric valleys of local minima can be seen at close and wide binary-lens separations, extending from $q\sim 2\times10^{-2}$ to 1. The best-fit model is unambiguously located in the wide separation regime, around ($s\simeq 3.28$, $q\simeq 0.73$).}
            \label{fig:chimap}
        \end{center}
    \end{figure*}

\section{Modelling}
    \label{sec:mod}

\subsection{Wide exploration of the parameter space}
\label{sec:gridsearch}

    The general shape of the OGLE-2015-BLG-0060 light curve, shown in Figure \ref{fig:lightcurve}, displays typical features associated with a binary-lens caustic-crossing event: a sharp rise in brightness as the source enters the caustic at HJD$\sim$2457146 (3 May), then a drop in brightness as the source traverses the interior of the caustic structure, followed by another sharp rise in brightness as the source exits the caustic at HJD$\sim$2457150 (7 May). The event then gradually returns to the standard Paczy\'{n}ski curve as the source moves further away from the caustic. The anomalous behaviour lasts for a total of $\sim$10 days, while the full duration of the event is $\gtrsim$120 days. 

    Strong binary microlensing features, such as those observed here, are subject to well-known model parameter partial degeneracies. These produce multiple local minima located within extended regions, rather than a single well-defined minimum \citep{kains2009}. Dedicated methods based on the observed features, such as the dates of the caustic crossings \citep{Cassan2008, cassan2010} may help in these cases to limit the extent of the region of parameter space to be explored. Regardless of the chosen fitting strategy, caustic-crossing events require exploring the morphology of the light curve for a wide range of binary-lens separations $s$ and mass ratios $q$, which can be visualised as a $\chi^2$-map in the $(s,q)$ plane. 

    At this stage of the analysis we wish to locate all regions of possible local minima without performing a computationally costly full parameter search. We assume that a simple binary-lens model including finite-source effects, but not parallax or orbital motion, (\textit{i.e.} a static binary-lens, straight line trajectory) can reproduce sufficiently well the gross features of the observed light curve. Besides parameters $s$ (projected binary-lens separation expressed in units of the angular Einstein radius of the lens $\theta_{\rm E}$) and mass ratio $q$, the others parameters of this model are: the impact parameter of the source $u_0$ expressed in $\theta_{\rm E}$ units; the characteristic duration of the event, $t_{\rm E}$ (or so-called Einstein time-scale), which is the time required for the source to cross the Einstein ring angular radius $\theta_{\rm E}$; the time of closest approach $t_0$ between the projected position of the source on the plane of the sky and the position of the centre of mass of the binary-lens; $\alpha$, the angle of the source trajectory with respect to the binary axis; finally $\rho=\theta_{\ast}/\theta_{\rm E}$, the angular source size expressed in $\theta_{\rm E}$ units. Finite-source effects become prominent when the source trajectory approaches or crosses a caustic, so it is mandatory to include them in the modelling even at this early stage.

    Hence, as a first step, we compute a high resolution $\chi^2(s, q)$ map, assuming a static binary-lens model including finite-source effects, using the Microlensing Search Map (MiSMap) algorithm. The underlying grid is uniformly spaced in $\log s$ and $\log q$, and samples 84 values of separation spanning values between $s=0.2$ and $5$, and 36 values in $q$ spanning values between $q=10^{-5}$ and $1$, for a total of 3024 grid points; for each grid point, we generate $10^{3}$ models to find the best-fit set of parameters. These models are derived from a refined library of pre-computed light curves for different values of $u_0$, $\alpha$, $t_\mathrm{E}$, $t_0$, and $\rho$. The resulting map is shown in Figure \ref{fig:chimap}. The blue regions mark the location of the valleys of local minima, while the red regions are zones of very unlikely sets of parameters. 
    The map clearly displays the classical degeneracy between models with $s=s_0$ or $s=1/s_0$ \citep{Dominik1999, ErdlSchneider1993}, which results in the two $\chi^2$ valleys highlighted in blue. Under our simple static binary-lens model assumption, we find that the global minimum is located in the upper part of the blue $\chi^2$ valley on the right (i.e. towards $q \sim 1$ and large values of $s$), extending approximately along the axis defined by $(s, q)\sim(1.5, 2\times10^{-2})$ to $(4, 1)$. In the regime we are considering (relatively large value of $s\sim2$ to $4$ and $q\sim 1$), the central and secondary caustics are of very similar shape and produce almost identical light curves, but which are still distinguishable in terms of $\chi^2$. The best-fit, marked with a cross, is clearly located at large separation, around $s \simeq 3.28$ and $q \simeq 0.73$. In the next Section, we shall see that this is already a very good estimate of the basic lens parameters, since adding parallax and/or orbital motion only slightly moves the best fit to larger values of $s$ and $q$ inside the upper part of the $\chi^2$ valley.

    Once this preliminary investigation of the parameter space is complete, we perform detailed modeling using as initial guesses the parameters found in the wide grid search (\textit{c.f.} Figure \ref{fig:chimap}). Besides the model parameters already mentioned, we include second-order parameters such as limb-darkening (Sec. \ref{sec:LLD}), parallax (Sec. \ref{sec:parallax}) and orbital motion between the two components of the lens (Sec. \ref{sec:OM}), and we discuss the final model in Sec. \ref{sec:bestmodel}. Our analysis uses the microlensing modeling software \textsc{muLAn} \citep[MICROlensing Analysis code,][]{muLAn}, which is an open-source code freely available online\footnote{https://github.com/muLAn-project/muLAn}. The software uses an Affine-Invariant Ensemble Sampler \citep{emcee2013, Goodman2010} to generate a multivariate proposal function while running several Markov Chain Monte Carlo (MCMC) chains for the set of parameters to be fitted. Where needed, some of the parameters are fitted within a predefined grid. We note that the results presented in this paper have been checked for consistency using the independently developed pyLIMA open-source package for microlensing modeling \citep{bachelet2017}.

\subsection{Source limb-darkening}
\label{sec:LLD}

    To take into account the limb-darkening of the source's extended surface, we model its surface brightness with a classical linear limb-darkening law, $S_\lambda(\vartheta) \propto 1 - \Gamma_\lambda (1 - 1.5\cos\vartheta)$, where $\vartheta$ is the angle between the line of sight toward the source star and the normal to the source surface, and $\Gamma_\lambda$ is the limb-darkening coefficient in pass-band $\lambda$. 

    Based on our estimate of the colour of the source star (which we describe in Sec. \ref{sec:sourcechar}), we adopt a temperature $T_{\rm{eff}}\sim 5250$ K \citep{ramirez2005} and use the \citet{claret2000} 
    tables to obtain the limb-darkening coefficients $u_\lambda$ for each pass-band. These values are then converted to linear limb-darkening model parameters through $\Gamma_\lambda= 2u_\lambda/(3-u_\lambda)$.
    This leads us to adopt $\Gamma_V=0.64$, $\Gamma_R=0.56$, $\Gamma_I=0.47$, $\Gamma_K = 0.25$, $\Gamma_{\mathrm{LIred}} = 0.47$ and $\Gamma_{\mathrm{MOAblue}} = 0.64$ that we keep fixed during the minimisation process (in fact, we find that fitting these parameters does not affect other best-fit parameters and $\chi^2$).
    
    Including limb-darkening and fitting the light curve by allowing $s$ and $q$ to be free parameters (as opposed to the initial $(s, q)$ grid search where they were fixed) leads to the best-fit static binary-lens model, whose parameters are given in Table \ref{tab:mod} (although the reported values and $\chi^2$ are those obtained after error bar re-scaling, \textit{c.f.} \ref{sec:bestmodel}).

\subsection{Parallax}
\label{sec:parallax}

    Given the long duration of the event ($\gtrsim$120 days), it is likely that the positional change of the observer caused by the orbital motion of the Earth around the Sun would have left a signature in the light curve. This so-called parallax effect causes the apparent lens-source motion to deviate from a rectilinear trajectory and manifests as a subtle long-term perturbation in the event light curve \citep[e.g.][]{gould1992,shin2012,jeong2015,street2016}. In order to model this effect, it is necessary to introduce two extra parameters, $\pi_{{\rm E},N}$ and $\pi_{{\rm E},E}$, representing the components of the parallax vector $\pivec_{\rm E}$ projected on the plane of the sky along the north and east equatorial axes respectively. With the inclusion of the parallax effect, we use the geocentric formalism of \cite{gould2004} which ensures that the parameters $t_0$, $u_0$ and $t_{\rm E}$ will be almost the same as when the event is fitted without parallax. In practice, the parallax is computed using real ephemerides rather than an approximation of the acceleration of the Earth around the Sun for the considered period of time. As a reference date, we choose $t_p=2457150$. 
    
    We inspect two classes of models that are expected to provide a similar (but not identical) fit to the data: one with $u_0>0$ and $\pi_{{\rm E},N}>0$, and the other with opposite signs $u_0<0$ and $\pi_{{\rm E},N}<0$. The resulting best-fit models have very similar $\chi^2$, though with a slight preference for $u_0<0$ as seen in Table \ref{tab:mod}. Compared to the static models that globally reproduce very well the main features of the light curve, we find that including parallax in the modeling substantially improves the $\chi^2$ of the fit (after error bar rescaling, c.f. Sec. \ref{sec:bestmodel}). 
    
    Parallax effects are thus clearly detected for this microlensing event. However, we now need to check whether orbital motion is also detected, since these two effects can be strongly degenerate.

\subsection{Orbital motion}
\label{sec:OM}

    Parallax is partly degenerate with the orbital motion of the binary-lens \citep{park2013,bachelet2012}. Orbital motion changes the shape of the caustics with time and, to first order approximation, can be modelled by introducing two more parameters that represent the rate of change of the normalised separation between the two lens components $ds/dt$ and the rate of change of the source trajectory angle relative to the caustics $d\alpha/dt$. 
    Given the possible degeneracy between parallax and orbital motion, we explore three classes of models: parallax alone, orbital motion alone and parallax plus orbital motion. 
    Our fits using a close binary model ($s<1$) always have much higher $\chi^2$ than those with a wide binary model $s>1$ ($\Delta\chi^2 > 4000$), so we discuss only the latter (for $q\sim 1$, which is the case here, we do not expect the classical $s-1/s$ degeneracy to occur). We also investigate the $u_0>0$ and $u_0<0$ degeneracy, resulting from the mirror-image symmetry of the source trajectory with respect to the binary-lens axis, which we are unable to break. As expected, the parameters of the event are almost identical for both cases. The calculations were repeated for different initial positions in parameter space to verify the uniqueness of the solution.

\begin{table*}
	\centering
	\caption{Best-fit microlensing parameters for the different competitive models. We choose the same reference date for parallax ($t_p$) and orbital motion ($t_b$), $t_p=t_b=2457150$ (HJD'=HJD-2450000). npar = number of fitted parameters.}
	\label{tab:mod}
	\begin{tabular}{lccccccr}
		\hline
		\hline
   	Parameter                 & static (npar=7) & \multicolumn{2}{c}{parallax (npar=9)}              & \multicolumn{2}{c}{parallax+orbital motion (npar=11)}\\
		\hline
                                  &                          &   $u_0>0$              &    $u_0<0$             &       $u_0>0$         &    $u_0<0$      \\  
        \hline
        $\chi^2$                  & 7738.59                  & 7589.02                & 7571.63                & 7383.51                & 7390.10        \\
        $t_0$ (HJD')              & 7030.72 $\pm$ 0.75       & 7038.65 $\pm$ 0.69     & 7039.06 $\pm$ 0.70     & 7046.22 $\pm$ 0.88     & 7046.07 $\pm$ 0.88 \\
        $u_0$                     & 0.779 $\pm$ 0.004        & 0.747 $\pm$ 0.003      & -0.746 $\pm$ 0.003     & 0.878 $\pm$ 0.013      & -0.831 $\pm$ 0.011 \\
        $t_{\rm E}$ (days)        & 73.90 $\pm$ 0.18         & 72.93 $\pm$ 0.18       & 72.81 $\pm$ 0.18       & 68.19 $\pm$ 0.55       & 66.703 $\pm$ 0.559 \\
       $s$                       & 3.487 $\pm$ 0.005         & 3.433 $\pm$ 0.005      & 3.430 $\pm$ 0.005      & 3.459 $\pm$ 0.006      & 3.470 $\pm$ 0.009 \\
       $q$                       & 0.817 $\pm$ 0.006         & 0.884 $\pm$ 0.006      & 0.885 $\pm$ 0.006      & 0.837 $\pm$ 0.007      & 0.834 $\pm$ 0.009 \\
       $\alpha$ (radians)        & 2.6952 $\pm$ 0.0006       & 2.6900 $\pm$ 0.0007    & -2.6894 $\pm$ 0.0007   & 2.622 $\pm$ 0.008      & -2.655 $\pm$ 0.007 \\
       $\rho$                    & 0.0046 $\pm$ $10^{-5}$    & 0.0048 $\pm$ $10^{-5}$ & 0.0047 $\pm$ $10^{-5}$ & 0.0047 $\pm$ $10^{-5}$ & 0.0047 $\pm$ $10^{-5}$ \\
       $\pi_{{\rm E},N}$         & --                        & 0.013 $\pm$ 0.002      & -0.018 $\pm$ 0.002     & 0.130 $\pm$ 0.012      & -0.087 $\pm$ 0.010 \\
       $\pi_{{\rm E},E}$         & --                        & -0.046 $\pm$ 0.003     & -0.044 $\pm$ 0.003     & -0.045 $\pm$ 0.005     & -0.064 $\pm$ 0.004 \\
       $ds/dt$ (yr$^{-1}$)       & --                        & --                     & --                     & 0.26 $\pm$ 0.09        & 0.63 $\pm$ 0.09 \\
       $d\alpha/dt$ (yr$^{-1}$)  & --                        & --                     & --                     & -0.28 $\pm$ 0.02       & 0.22 $\pm$ 0.02 \\
        \hline
    \end{tabular}
\end{table*}

\subsection{Discussion and best-fit model}
\label{sec:bestmodel}

    The final step is the refinement of the model by adjusting the uncertainties of each data set and refitting. The data sets used in this analysis are obtained from different telescopes and instruments with notable differences in their photometric precision and the measurement errors are often underestimated. We therefore normalise the reported flux uncertainties of the $i$th data set using the expression $e_i = f_i (\sigma_0^2 + \sigma_i^2)^{1/2}$, where $f_i$ is a scale factor, $\sigma_0$ are the originally reported uncertainties and $\sigma_i$ is an additive uncertainty term for each data set $i$ \citep{2012ApJ...755..102Y}. Thus, the error-bars are adjusted so that the $\chi^2$ per degree of freedom ($\chi^2$/dof) of each data set relative to the model is one. The model is then recomputed.

\begin{figure*}
	    \includegraphics[width=2\columnwidth]{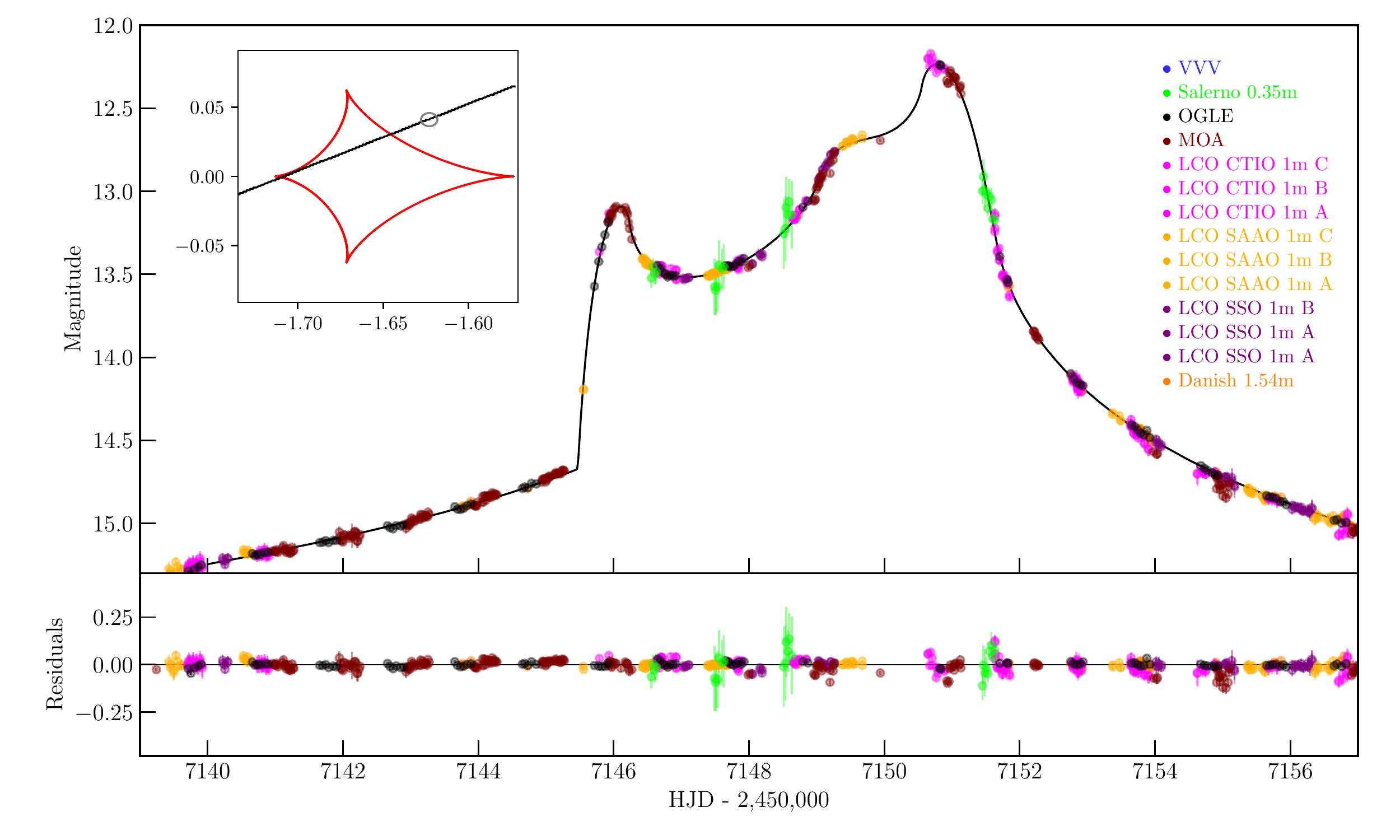}
        \caption{Enlarged view around the peak of the light curve of microlensing event OGLE-2015-BLG-0060 highlighting the anomalous structure and showing the best-fit binary model including parallax. The legend on the right of the figure lists the contributing telescopes. The inset on the top left displays the caustic pattern for this event, while the black line indicates the source trajectory. The source size (in units of $\theta_E$) is represented by the small grey circle. The points of entry and exit are associated with the most highly magnified features in the light curve at HJD$\sim$2457146 and HJD$\sim$2457151 respectively.}
        \label{fig:lc_zoom}
    \end{figure*}

    The results of the modelling runs are summarised in Table \ref{tab:mod}. The reported uncertainties for each parameter correspond to the size of the one-sigma contours of the parameter error distributions generated by the MCMC chains. 
    The best-fit orbital-motion-only model ($u_0<0$) has a $\chi^2=$7621.61 and is disfavoured compared to the best-fit parallax-only model ($u_0<0$) which has a $\chi^2=$7571.63. The $\chi^2$ improves by $\sim 180$ when orbital motion is considered together with parallax. However, as we discuss in the next Section, these solutions result in unbound systems and are therefore rejected. The best-fit binary-lens model with parallax is presented in Figure \ref{fig:lightcurve}, superposed on the data. Figure \ref{fig:lc_zoom} shows an enlarged view of the region around the peak, where the perturbations are prominent. The source trajectory with respect to the caustic structure is shown as an inset. It crosses the caustic structure twice, causing a substantial increase in magnification at the entry and exit points. Follow-up observations cover all the critical features present in the light curve.

\subsection{Survey vs. follow up}
\label{sec:surveyfollowup}
    Our previously described fits include survey as well as follow-up data. To test to what extent each data set can constrain the parameters without the other, we perform three separate fits starting from a full exploration of the parameter space: OGLE+MOA, OGLE-only and follow-up only. Table \ref{tab:sets} shows the results of these fits. For the survey data, in analogy to the all-data fit, the model including parallax and orbital motion gives the best fit ($\chi^2 \sim$ 6382), followed by the parallax-only-model ($\chi^2 \sim$ 6557). The main microlensing fit parameters are in good agreement with our results from the all-data fits. However, the parallax vector, and especially the north component, $\pi_{{\rm E},N}$,  shows strong variation between the different fits.
    
  In the OGLE data an additional feature can be identified prior to the main event: A broad low magnification peak observed at the end of the 2014 observing season at HJD$\sim$2456920, as alerted by V. Bozza (see section~\ref{detanom}). This feature can not be seen in the MOA data and is not covered by the follow-up data sets. The scatter and uncertainties in the MOA data are larger than the low amplitude of this feature so they cannot constrain it. To assess how this "bump" influences the parallax values we perform a fit using OGLE data alone. The parallax signal is now more strongly detected ($\pi_{{\rm E},N}$ = -0.0215$\pm$ 0.0038). Given that parallax is a long-term effect, this is expected. Next, we perform an all-data fit but fix the parallax to the value determined by the OGLE-only fit. The result of this fit has a $\chi ^2$ that is worse by $\sim 18$ compared to our best fit using all available data but, most notably, this difference is fully attributed to a failure of this model to match the $\sim$20 points of the first peak of the anomaly (HJD$\sim$2457146) during the caustic entry. 
  This suggests that OGLE data alone are not sufficient to fully constrain the parallax, and that the full data set that covers the structure of the peak anomaly remarkably well is required. We also note that, in contrast to the all data-fit, the OGLE-only fit results in a bound system for a source located at the distance of the RC. The inclusion of follow-up data is therefore crucial in deciding between competing solutions. 

The fit with follow-up data alone shows similar results. The best fit parameters are close to our fit using all data, with the exception of the parallax signal which is only weakly detected. From this we conclude that both survey and follow-up data are needed to reliably constrain the parallax.
\begin{table*}
\centering
\caption{Best-fit parameters for the parallax-only models ($u_0<0$) including different data sets. Since the data sets contain different numbers of data points, the reduced-$\chi^2$ is reported ($\chi ^2$/dof, dof = degrees of freedom). }
\label{tab:sets}
\begin{tabular}{lccc}
\hline
\hline
 Parameter   	     	  & OGLE + MOA         	   & OGLE               	& follow-up          \\
\hline
 $\chi ^2$/dof 	   		  & 0.878           	   & 0.883           		& 0.779          \\
 $t_0 $ (HJD')     		  & 7037.92 $\pm$ 0.76     & 7048.54$\pm$ 1.31   	& 7045.13 $\pm$ 0.95   \\
 $u_0$             		  & -0.756 $\pm$ 0.004     & -0.709 $\pm$ 0.008 	& -0.720 $\pm$ 0.006 \\
 $t_{\rm E}$ (days)		  & 72.98 $\pm$ 0.19       & 70.28 $\pm$ 0.31 	    & 68.98 $\pm$ 0.30 \\
 s             	   		  & 3.433 $\pm$ 0.006      & 3.353 $\pm$ 0.010  	& 3.374 $\pm$ 0.008  \\
 q                 		  & 0.870 $\pm$ 0.007      & 0.939 $\pm$ 0.012  	& 0.867 $\pm$ 0.011  \\
 $\alpha$ (radians)		  & -2.687 $\pm$ 0.001     & -2.689 $\pm$ 0.002 	& -2.703 $\pm$ 0.002 \\
 $\rho$            		  & 0.0048 $\pm$ $10^{-5}$ & 0.0051 $\pm$ $10^{-5}$ & 0.0050 $\pm$ $10^{-5}$ \\
 $\pi_{{\rm E},N}$ 		  & -0.004 $\pm$ 0.004     & -0.022 $\pm$ 0.004 	& -0.003 $\pm$ 0.003 \\
 $\pi_{{\rm E},E}$ 		  & -0.049 $\pm$ 0.003     & -0.055 $\pm$ 0.005 	& -0.057 $\pm$ 0.006 \\
 $ds/dt$ (yr$^{-1}$) 	  & --				   	   & --				      	& --  \\
 $d\alpha/dt$ (yr$^{-1}$) & --				   	   & --				    	& --  \\
\hline
\end{tabular}
\end{table*}    
\section{Physical parameters}
\label{sec:phy}

    \begin{figure}
        \begin{center}
            \includegraphics[width=1\columnwidth]{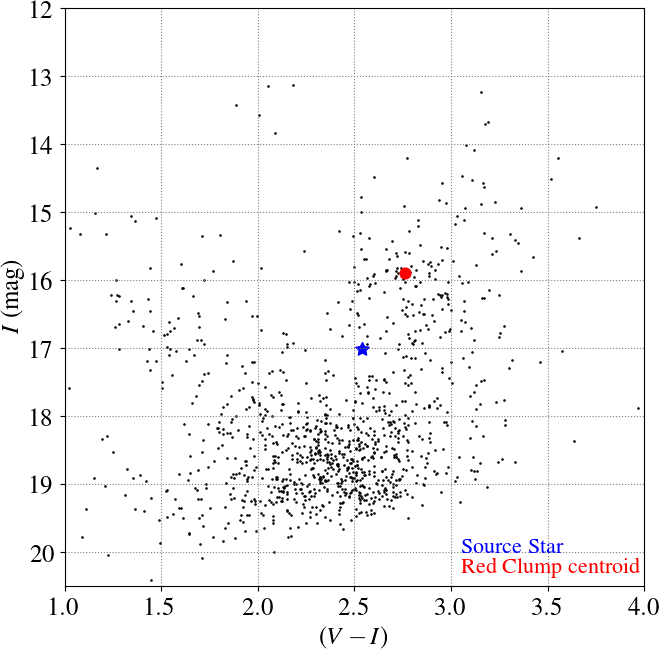}
            \caption{$(V-I, I)$ CMD of stars from the OGLE-IV catalog within $90\arcsec$ from the location of OGLE-2015-BLG-0060, not corrected for interstellar extinction. The red point corresponds to the location of the Red Clump and the blue star to the location of the source.}
            \label{fig:cmd}
        \end{center}
    \end{figure}

\subsection{Source star characterisation}
\label{sec:sourcechar}

    To estimate the angular source radius, we use OGLE-IV $I$ and $V$-band observations of stars within a $90\arcsec$ radius centred on the microlensing target to generate the $(V-I,I)$ colour-magnitude diagram (CMD) shown in Figure~\ref{fig:cmd}. We then identify the centroid of the red clump (RC) at $(V-I,I)_\mathrm{RC,OGLE} = (2.76\pm 0.02, 15.90 \pm 0.05)$. The unblended instrumental colour and magnitude of the source star is evaluated during the model fit: $(V-I,I)_\mathrm{S,OGLE} = (2.54\pm 0.02, 17.02 \pm 0.01)$. This yields an offset of $\Delta(V-I,I)_\mathrm{OGLE} = (V-I,I)_\mathrm{S,OGLE} - (V-I,I)_\mathrm{RC,OGLE} = (-0.22\pm 0.03, 1.12\pm 0.05)$. To account for OGLE's non-standard $V$ band, the $\Delta(V-I)_\mathrm{OGLE}$ value needs to be multiplied with 0.93 \citep{Udalski2015} to bring it to the standard Johnson-Cousins (JC) system, yielding $\Delta(V-I)_\mathrm{JC} = -0.21\pm 0.03$. The intrinsic mean dereddened colour and apparent magnitude of the RC (at the coordinates of the microlensing event) are $(V-I)_\mathrm{RC,0} = 1.06\pm 0.03$ \citep{Bensby2013,nataf2013} and $(I)_\mathrm{RC,0} = 14.36\pm 0.02$ \citep{nataf2016} respectively.

    The distance to the RC can be derived from the measurement of the distance to the Galactic centre (GC) \citep{nataf2016}, $D_\mathrm{GC}=8.33\ \mathrm{kpc}$, by computing
    \begin{equation}
        D_\mathrm{RC} = \frac{D_\mathrm{GC} \sin{\phi}}{\cos(b)\sin{(l+\phi)}},
    \end{equation}
    where $\phi=40^\circ$ is the angle between the major axis of the Galactic bulge and the line of sight from the Sun. For OGLE-2015-BLG-0060, we find the RC to be on the close side of the bar at a distance of $D_\mathrm{RC}=7.9$ kpc, corresponding to a distance modulus of $\mu=14.5\ \mathrm{mag}$.

    Assuming that the reddening towards the microlensing source is the same as towards the RC and that the distance to the source is the same as the distance to the RC, the intrinsic (dereddened) colour and magnitude of the source can be estimated: $(V-I)_\mathrm{S,0} = \Delta(V-I)_\mathrm{JC} + (V-I)_\mathrm{RC,0} = 0.85\pm 0.04$ and $I_\mathrm{S,0} = I_\mathrm{RC,0} - I_\mathrm{RC,OGLE} + I_\mathrm{S,OGLE} + 0.00168 = 15.48\pm 0.06$. Therefore the source star is most probably a G-type sub-giant. From the dereddened colour and magnitude of the source, we can estimate the angular source radius $\theta_*$ \citep{kervella2008} using
    \begin{equation}
        \log\left(\thS\right)=3.198-0.2 I_\mathrm{s,0}+0.4895 (V-I)_{\mathrm{s},0}-0.0657 (V-I)^2_{\mathrm{s},0},
    \end{equation}
    where the angular radius is given in $\mu$as and the uncertainty of the relation is 0.0238. This yields the angular source size, $\theta_* = 2.96\pm 0.36 \mu$as.
    
    We then proceed to evaluate the angular Einstein radius, $\theta_{\rm E} = \theta_*/\rho = 0.62\pm 0.08$ mas, and the (geocentric) lens-source relative proper motion, $\mu_{\rm geo}=\theta_{\rm E}/t_{\rm E} = 2.91\pm 0.35$ mas yr$^{-1}$.

\subsection{Physical parameter estimation}
    The mass and distance to the lens are determined by
    \begin{equation}
        M_{\rm tot}=\frac{\theta_{\rm E}}{\kappa \pi_{\rm E}}; \qquad 
        D_{\rm L}=\frac{\rm AU}{\pi_{\rm E}\theta_{\rm E}+\pi_{\rm S}},
    \end{equation}
    where $\kappa=4G/(c^2{\rm AU})$ and $\pi_{\rm S} = {\rm (AU)}/D_{\rm S}$ is the parallax of the source star \citep{gould1992}. To determine these quantities, we need $\pi_{\rm E}$ and $\theta_{\rm E}$. The value of $\pi_{\rm E}$ is estimated from the model fit, whereas $\theta_{\rm E}=\theta_{\ast}/\rho_{\ast}$ depends on the angular radius of the source star, $\theta_{\ast}$, and the normalised source radius, $\rho_{\ast}$, which is also returned from modelling (see Table~\ref{tab:mod}). Therefore, to get the value of $\theta_{\rm E}$, we needed first to estimate $\theta_{\ast}$.

    Our analysis indicates that the lens is a binary system comprised of two stars with almost equal mass ($q$=0.86-0.9). The physical parameters of the system are presented in Table \ref{tab:phy} based on the best binary-lens model with parallax but no orbital motion. 
    \begin{table}
	    \centering
	    \caption{Physical parameters}
	    \label{tab:phy}
	    \begin{tabular}{ll}
	       \hline
	   	   \hline
	    	Parameter                                    & Value                         \\ 
	    	\hline
            Mass of lens star \#1 ($M_{\star1}$)         & 0.87 $\pm$ 0.12 $M_{\odot}$   \\ 
            Mass of lens star \#2 ($M_{\star2}$)         & 0.77 $\pm$ 0.11 $M_{\odot}$   \\ 
            Distance to the lens ($D_{\rm L}$)           & 6.41 $\pm$ 0.14 kpc           \\ 
            Projected star-star separation ($d_\perp$)   & 13.85 $\pm$ 0.16 AU            \\ 
            Einstein radius ($\theta_{\rm E}$)           & 0.62 $\pm$ 0.08 mas           \\ 
            Geocentric proper motion ($\mu_{\rm geo}$)   &  3.16$\pm$ 0.39 mas yr$^{-1}$ \\
            \hline
        \end{tabular}
    \end{table}

    The two models with parallax and orbital motion produce better fits in terms of their corresponding $\chi^2$ values, but when we evaluate the ratio of kinetic to potential energy \citep{2018ApJ...853...70U}
    \begin{equation}
        \left(\frac{KE}{PE}\right)_\perp = \frac{(d_\perp/AU)^3}{8\pi^2(M_L/M_\odot)} \left[\left(\frac{1}{s}\frac{ds}{dt}\right)^2 + \left(\frac{d\alpha}{dt}\right)^2\right],
    \end{equation}
    we find that $(KE/PE)_\perp > 1.5$ for both of them, which results in unbound systems\footnote{The ratio of the kinetic to potential energy should be less than 1 for the system to be bound.}. Furthermore, the large projected distance between the two components ($\sim14$ AU) implies an orbital period of $\sim$36 years. Such a long period, when compared to the event timescale of $\sim70$ days, suggests that any orbital motion effects would be negligible. 
    We therefore conclude that the improvement in the $\chi^2$ from the inclusion of the orbital motion is not due to a physical effect, and is most likely caused by long-term systematics in the data.

    Adopting the parameters of the model including parallax but no orbital motion, the distance to the lens is $D_{\rm L} = 6.41\pm 0.14$\ kpc, in the direction of the Galactic Bulge. The two components have masses $M_{\star1} = 0.87\pm 0.12$\ $M_{\odot}$ and  $M_{\star2} = 0.77\pm 0.11$\ $M_{\odot}$ respectively. The projected separation between them is $d_\perp = 13.85\pm 0.16$\ AU. 
    
    \subsection{Close source interpretation}
    What if the source were closer? Source distances $\leq$ 2.85 kpc can lead to bound solutions. Even though the lensing probability for such nearby sources is extremely small, we explore this possibility next for the sake of completeness. We identified a star at the coordinates of the microlensing event in the Gaia DR2 catalogue \citep{2016A&A...595A...1G,2018A&A...616A...1G} and used the distances from \citet{2018AJ....156...58B} to derive a distance for the source of $D_S = 2.67^{+3.58}_{-1.22}$kpc at 68\% confidence level\footnote{Note that the Gaia measured parallax is consistent with infinite distance at 2$\sigma$.}). Using the VPHAS+ DR2 catalogue \citep{2014MNRAS.440.2036D}, we generated a colour-colour diagram using a search radius or 60 arcsec around the coordinates of the event. The resulting distribution was compared with the atlas of synthetic spectra of \citet{1985ApJS...59...33P} and the location of the source on the diagram implied that it is likely a G-type Main Sequence (MS) star. To estimate the reddening at the assumed source distance of 2.67 kpc, we used the Python {\it dustmaps} package \citep{2018JOSS....3..695M}, which assumes the extinction law derived in \citet{2016ApJ...821...78S}. We then transformed to different passbands using \citet{2011ApJ...737..103S} and derived $A_I = 0.4\pm 0.20$. Applying this correction and repeating the calculations in Section~\ref{sec:sourcechar} we obtained $(V-I)_\mathrm{S,0} = 2.13\pm 0.09$ and $I_\mathrm{S,0} = 16.52\pm 0.21$, which implies a larger angular source radius $\thS = 4.351\pm 0.519$ $\mu$as. The derived physical parameters of the system then become $M_{\star1} = 0.58\pm 0.08$\ $M_{\odot}$, $M_{\star2} = 0.48\pm 0.07$\ $M_{\odot}$, $d_\perp = 6.77^{+1.35}_{-0.49}$ AU, leading to a bound system. 

\section{Conclusions}
\label{sec:con} 

    We analysed the binary microlensing event OGLE-2015-BLG-0060. The caustic-crossing features of the light curve were sampled intensively with automated follow-up observations from the robotic telescopes of the Las Cumbres Observatory. The trajectory of the source star crosses the central caustic structure twice, entering at HJD$\sim$2457146 (3 May) and exiting at HJD$\sim$2457150 (7 May). The light curve does not display the typical ``U''-shape associated with binary-lenses, but displays a ``bump'' between the entry and exit points, which is associated with the source trajectory approaching a cusp. 
    We found that considering the parallax is necessary to explain the morphology of the light curve. The two components of the binary-lens have masses $M_{\star1}  = 0.87 M_{\odot}$ and $M_{\star2}  = 0.77 M_{\odot}$, and a projected separation of $d_{\perp} = 13.85$ AU. The effect of orbital motion is negligible because of the wide separation between the lensing components, which implies a long orbital period for the binary (P$\sim$40 years compared to $t_E\sim$77 days). The distance to the lensing system is 6.4 kpc.
    We are unable to break the ecliptic degeneracy, i.e. the degeneracy caused by the mirror symmetry between the source trajectories with $u_0>0$ and $u_0<0$ with respect to the binary axis. 
    This degeneracy does not affect our estimate of the physical parameters of the lensing system since the underlying model parameters have similar values.
    Finally, we considered possible alternative interpretations of the event under the assumption of a nearby source star.
    
    This work demonstrates that timely reactive observations from robotic telescopes are already capable of achieving excellent automatic coverage of anomalous light curve features. However, to place meaningful constraints on the physical parameters of the lens, observations on the wings and baseline of the light curve are essential.

\section*{Acknowledgements}

    AC acknowledges financial support from Universit{\'e} Pierre et Marie Curie under grant {\'E}mergence@Sorbonne Universit{\'e}s 2016. This work was granted access to the HPC resources of the HPCaVe at UPMC-Sorbonne Universit{\'e}. SM was supported by the National Science Foundation of China (Grant No. 11333003, 11390372 to SM). KH acknowledges support from STFC grant ST/M001296/1. YT, JW acknowledge the support of the DFG priority program SPP 1992 "Exploring the Diversity of Extrasolar Planets (WA 1047/11-1). The OGLE project has received funding from the National Science Centre, Poland, grant MAESTRO 2014/14/A/ST9/00121 to AU. Work by RAS and EB was supported by NASA grant NNX15AC97G. This work made use of observations from the LCOGT network, which includes three SUPAscopes owned by the University of St. Andrews. The MOA project is supported by JSPS KAKENHI Grant Number JSPS24253004, JSPS26247023, JSPS23340064, JSPS15H00781, and JP16H06287. The work  by  CR  was supported  by  an  appointment  to the NASA Postdoctoral Program  at the  Goddard  Space  Flight  Center, administered by USRA through a contract with NASA. This work has made use of data from the European Space Agency (ESA) mission {\it Gaia} (\url{https://www.cosmos.esa.int/gaia}), processed by the {\it Gaia} Data Processing and Analysis Consortium (DPAC, \url{https://www.cosmos.esa.int/web/gaia/dpac/consortium}). Funding for the DPAC has been provided by national institutions, in particular the institutions participating in the {\it Gaia} Multilateral Agreement. Based on data products from observations made with ESO Telescopes at the La Silla Paranal Observatory under public survey programme ID, 177.D-3023\\

\noindent 
$^{1}$Zentrum f{\"u}r Astronomie der Universit{\"a}t Heidelberg, Astronomisches Rechen-Institut, M{\"o}nchhofstr. 12-14, 69120 Heidelberg, Germany\\
$^{2}$Institut d'Astrophysique de Paris, Sorbonne Universit\'e, CNRS, UMR 7095, 98 bis bd Arago, 75014 Paris, France\\
$^{3}$Las Cumbres Observatory Global Telescope Network, 6740 Cortona Drive, suite 102, Goleta, CA 93117, USA\\
$^{4}$School of Physical Sciences, University of Tasmania, Private Bag 37
Hobart, Tasmania 7001 Australia\\
$^{R13}$Planetary and Space Sciences, Department of Physical Sciences, The Open University, Milton Keynes, MK7 6AA, UK\\
$^{R4}$Astrophysics Research Institute, Liverpool John Moores University, Liverpool CH41 1LD, UK\\
$^{R3}$SUPA, School of Physics \& Astronomy, University of St Andrews, North Haugh, St Andrews KY16 9SS, UK\\
$^{R10}$Dipartimento di Fisica "E.R. Caianiello", Universit{\`a} di Salerno, Via Giovanni Paolo II 132, 84084 Fisciano, Italy\\
$^{R15}$National Astronomical Observatories, Chinese Academy of Sciences, 20A Datun Road, Chaoyang District, Beijing 100012, China\\
$^{R16}$Physics Department and Tsinghua Centre for Astrophysics, Tsinghua University, Beijing 100084, China\\
$^{R17}$Jodrell Bank, School of Physics and Astronomy, The University of Manchester, Oxford Road, Manchester M13 9PL, UK\\
$^{R18}$New York University Abu Dhabi, PO Box 129188, Saadiyat Island, Abu Dhabi, UAE\\
$^{R19}$Institute for Astronomy, University of Edinburgh, Royal Observatory, Edinburgh EH9 3HJ, UK\\
$^{R20}$International Space Science Institute (ISSI), Hallerstrasse 6, 3012 Bern, Switzerland\\
$^{R8}$South African Astronomical Observatory, PO Box 9, Observatory 7935, South Africa\\
$^{S2}$Istituto Nazionale di Fisica Nucleare, Sezione di Napoli, Via Cintia, 80126 Napoli, Italy\\
$^{S3}$IPAC, Mail Code 100-22, Caltech, 1200 E. California Blvd., Pasadena, CA 91125, USA\\
$^{S4}$INAF-Osservatorio Astronomico di Capodimonte, Salita Moiariello 16, 80131, Napoli, Italy\\
$^{S5}$Istituto Internazionale per gli Alti Studi Scientifici (IIASS), Via G. Pellegrino 19, 84019 Vietri sul Mare (SA), Italy\\
$^{O1}$Warsaw University Observatory, Al.~Ujazdowskie~4, 00-478~Warszawa, Poland\\
$^{O2}$Department of Astronomy, Ohio State University, 140 W. 18th Ave., Columbus, OH  43210, USA\\
$^{M1}${Department of Earth and Space Science, Graduate School of Science, Osaka University, Toyonaka, Osaka 560-0043, Japan}\\
$^{M2}${Institute for Space-Earth Environmental Research, Nagoya University, Nagoya 464-8601, Japan}\\
$^{M3}${Code 667, NASA Goddard Space Flight Center, Greenbelt, MD 20771, USA}\\
$^{M4}${Department of Physics, University of Auckland, Private Bag 92019, Auckland, New Zealand}\\
$^{M5}${Okayama Observatory, National Astronomical Observatory of Japan, 3037-5 Honjo, Kamogata, Asakuchi, Okayama 719-0232, Japan}\\
$^{M6}${Institute of Natural and Mathematical Sciences, Massey University, Auckland 0745, New Zealand}\\
$^{M7}${Nagano National College of Technology, Nagano 381-8550, Japan}\\
$^{M8}${Tokyo Metropolitan College of Aeronautics, Tokyo 116-8523, Japan}\\
$^{M9}${School of Chemical and Physical Sciences, Victoria University, Wellington, New Zealand}\\
$^{M10}${University of Canterbury Mt.\ John Observatory, P.O. Box 56, Lake Tekapo 8770, New Zealand}\\
$^{M11}${Department of Physics, Faculty of Science, Kyoto Sangyo University, 603-8555 Kyoto, Japan}\\
$^{M12}${Department of Astronomy, University of Maryland, College Park, MD 20742, USA}\\
$^{P1}${Niels Bohr Institute \& Centre for Star and Planet Formation, University of Copenhagen {\O}\\ster Voldgade 5, 1350 - Copenhagen, Denmark}\\
$^{P3}${Max Planck Institute for Solar System Research, Max-Planck-Str. 2, 37191 Katlenburg-Lindau, Germany}\\
$^{P4}${Qatar Environment and Energy Research Institute (QEERI), HBKU, Qatar Foundation, Doha, Qatar}\\
$^{P5}${Qatar Foundation, P.O. Box 5825, Doha, Qatar}\\
$^{P10}${Max Planck Institute for Astronomy, K{\"o}nigstuhl 17, 69117 Heidelberg, Germany}\\
$^{P11}${Finnish Centre for Astronomy with ESO (FINCA), V{\"a}is{\"a}l{\"a}ntie 20, FI-21500 Piikki{\"o}, Finland}\\
$^{P12}${European Southern Observatory, Karl-Schwarzschild Stra\ss{}e 2, 85748 Garching bei M\"{u}nchen, Germany}\\
$^{P14}${Astrophysics Group, Keele University, Staffordshire, ST5 5BG, UK}\\
$^{P15}${Unidad de Astronom{\'{\i}}a, Fac. de Ciencias B{\'a}sicas, Universidad de Antofagasta, Avda. U. de Antofagasta 02800, Antofagasta, Chile}\\
$^{P16}${Korea Astronomy \& Space Science Institute, 776 Daedukdae-ro, Yuseong-gu, 305-348 Daejeon, Republic of Korea}\\
$^{P17}${Universit{\"a}t Hamburg, Faculty of Mathematics, Informatics and Natural Sciences, Department of Earth Sciences, Meteorological Institute, Bundesstra\ss{}e 55, 20146 Hamburg, Germany}\\
$^{P18}${Stellar Astrophysics Centre, Department of Physics and Astronomy, Aarhus University, Ny Munkegade 120, DK-8000 Aarhus C, Denmark}\\
$^{P19}${Istituto Internazionale per gli Alti Studi Scientifici (IIASS), Via G. Pellegrino 19, 84019 Vietri sul Mare (SA), Italy}\\
$^{P21}${Dark Cosmology Centre, Niels Bohr Institute, University of Copenhagen, Juliane Maries Vej 30, 2100 - Copenhagen {\O}, Denmark}\\
$^{P22}${Department of Physics, Sharif University of Technology, PO Box 11155-9161 Tehran, Iran}\\
$^{P23}${Centre for Electronic Imaging, Department of Physical Sciences, The Open University, Milton Keynes, MK7 6AA, UK}\\
$^{P24}${Yunnan Observatories, Chinese Academy of Sciences, Kunming 650011, China}\\
$^{P25}${Institut d'Astrophysique et de G\'eophysique, All\'ee du 6 Ao\^ut 17, Sart Tilman, B\^at. B5c, 4000 Li\`ege, Belgium}\\
$^{P26}${Instituto de Astrof\'isica, Facultad de F\'isica, Pontificia Universidad Cat\'olica de Chile, Av. Vicu\~na Mackenna 4860, 7820436 Macul, Santiago, Chile}\\
$^{P27}${CITEUC -- Centre for Earth and Space Science Research of the University of Coimbra, Observat\'orio Astron\'omico da Universidade de Coimbra, 3030-004 Coimbra, Portugal}\\
$^{P28}${Max Planck Institute for Astronomy, K{\"o}nigstuhl 17, 69117 Heidelberg, Germany}\\
$^{P29}${Department of Physics, University of Rome Tor Vergata, Via della Ricerca Scientifica 1, I-00133Roma, Italy}\\
$^{P30}${INAF -- Astrophysical Observatory of Turin, Via Osservatorio 20, I-10025 -- Pino Torinese, Italy}
\bibliographystyle{mnras}
\input{bibliography.bbl}


\appendix
\section{Additional material}
\label{appx:bump}
\begin{figure*}
\begin{center}
	\includegraphics[width=2\columnwidth]{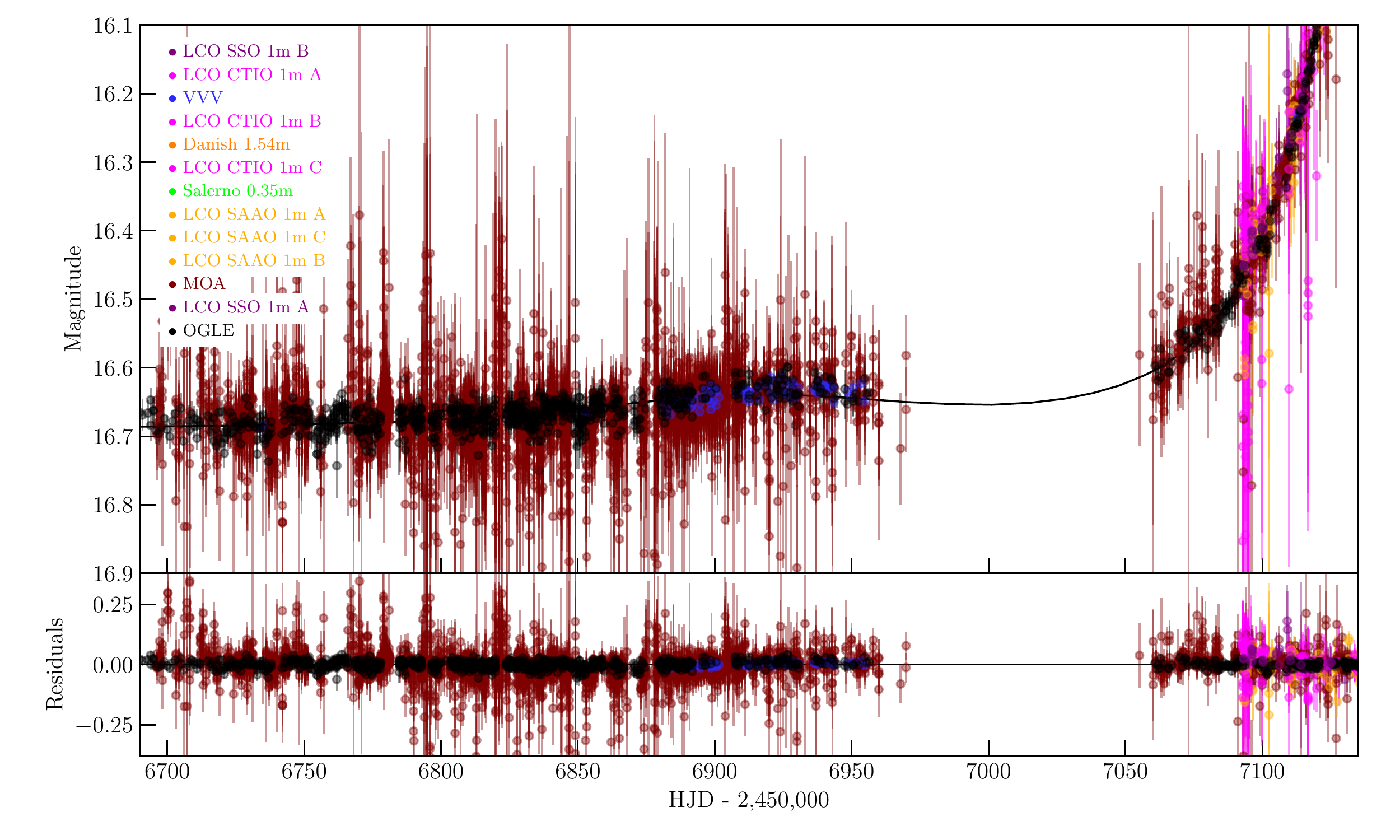}
    \caption{Enlarged view of the light curve of microlensing event OGLE-2015-BLG-0060, centred on the broad peak that was observed at the end of the 2014 season, prior to the main event. The legend on the left lists the observations used to model this event. The solid black curve represents our best-fit model to the data.}
    \label{fig:bump}
\end{center}
\end{figure*}





\bsp	
\label{lastpage}
\end{document}